%% file: neurips_2026.tex
\documentclass{article}

\PassOptionsToPackage{numbers, compress}{natbib}

\usepackage[main, final]{neurips_2026}

\usepackage[utf8]{inputenc} %
\usepackage[T1]{fontenc}    %
\usepackage{hyperref}       %
\usepackage{url}            %
\usepackage{booktabs}       %
\usepackage{amsfonts}       %
\usepackage{nicefrac}       %
\usepackage{microtype}      %
\usepackage{xcolor}         %
\usepackage{amsmath,amssymb,amsthm}
\usepackage{enumitem}       %

\theoremstyle{definition}

\newtheorem{assumption}{Assumption}[section]
\newtheorem{proposition}{Proposition}[section]

\theoremstyle{plain}
\newtheorem{lemma}{Lemma}[section]
\newtheorem{theorem}{Theorem}[section]

\theoremstyle{remark}

\usepackage{tcolorbox}
\usepackage{multirow}
\usepackage{colortbl} 
\usepackage{array}
\usepackage{rotating}
\usepackage[export]{adjustbox}
\usepackage{subfigure}
\usepackage{graphicx}
\usepackage{algorithm}
\usepackage{algpseudocode}

\title{Phase-wise MLLM Tuning for Multi-framework WebUI Code Generation}

\author{%
  Haoran Ma$^{1*}$ \quad
  Linxiao Li$^{2*}$ \quad
  Chenyue Wang$^{2}$ \quad
  Haochen Sui$^{3}$ \quad
  Jiechao Gao$^{4\dagger}$ \\[4pt]
  $^{1}$Harvard University \quad
  $^{2}$The University of Sydney \\
  $^{3}$University of Michigan \quad
  $^{4}$Stanford University
}

\begin{document}

\maketitle
{
\renewcommand{\thefootnote}{\fnsymbol{footnote}}
\footnotetext[1]{Equal contribution.}
\footnotetext[2]{Corresponding author: \texttt{jiechao@stanford.edu}.}
}

\begin{abstract}
Multimodal large language models (MLLMs) achieve strong performance in translating WebUI screenshots into HTML/CSS. 
However, when they faced multiple frameworks (React/Vue/Angular), they often suffer from negative transfer that leads to compilation failures. 
These failures commonly stem from mixing framework specific syntax and violating framework constraints.
This paper studies supervised fine-tuning for multi-framework WebUI code generation and proposes a phase-wise MLLM tuning method driven by compatibility and heterogeneity signals across frameworks. 
We first apply a normalization mapping that preserves each framework's native syntax while suppressing noise from variable naming and literal content. 
We then estimate cross framework heterogeneity and compatibility offline, and automatically derive training groups and their ordering by maximizing a phased objective. 
Finally, we perform phase-wise adapter tuning to reduce cross framework interference. 
Experiments on multiple benchmarks show that our proposed method improves both generation quality and compilation success rate compared with baselines.  
\end{abstract}

\section{Introduction}
Generating executable front-end code from WebUI screenshots has become an important objective in automated software engineering \cite{si2025design2code}. Automating this process reduces the manual effort required to implement user interfaces and shortens the path from visual design to deployable code \cite{zhou2025declarui}. Early systems relied on computer vision pipelines that extracted structural elements from interface images and then used sequence models to emit HTML/CSS \cite{nguyen2015reverse, beltramelli2018pix2code, chen2018ui, xu2021image2emmet}. With the emergence of MLLMs, recent work has shifted toward end-to-end approaches that produce executable code directly from screenshots by jointly modeling visual inputs and code representations \cite{laurenccon2024unlocking, yun2024web2code, gui2025uicopilot}.

Despite this progress, most existing studies assume that the generated code targets HTML/CSS \cite{yun2024web2code, gui2025webcode2m}. In practical development, however, WebUI projects often need to support several widely used frameworks, including React, Vue, and Angular, which differ in syntax conventions, component organization, and engineering constraints \cite{xiao2025designbench}. When a model is trained jointly across such heterogeneous targets, or transferred from a model trained on a single framework, shared parameters receive supervision from distributions that favor different output patterns \cite{kollias2021distribution}. This can lead to cross-framework negative transfer \cite{yu2020gradient}, where generated code mixes syntax from different frameworks or violates framework-specific rules, ultimately causing compilation failures \cite{wu2024uicoder}.

The root of this difficulty lies not only in the output distributions but also in the training dynamics induced by each framework. Heterogeneous target distributions produce gradient directions on the shared parameters that are not necessarily aligned across frameworks \cite{sener2018multi, liu2021conflict}. When these signals are aggregated without regard to their relationships, shared updates may reflect inconsistent objectives, resulting in gradient interference and unstable optimization \cite{standley2020tasks, fifty2021efficiently}. This raises a natural question: In multi-framework WebUI code generation, can training strategies explicitly account for differences and compatibility among frameworks in order to reduce cross-framework interference?

To investigate this question, this paper studies supervised fine-tuning for multi-framework WebUI code generation and introduces a phase-based training strategy for MLLMs. The method first normalizes each framework's code to suppress superficial variation such as variable names and literal content. It then characterizes each pair of frameworks through two signals—a heterogeneity measure over normalized code distributions and a compatibility measure based on pre-gradients at the shared-adapter insertion point—and uses these signals to partition frameworks into training phases and order them. During training, the model combines a shared adapter with framework-specific adapters and updates them phase by phase, so that frameworks with more consistent optimization behavior are learned together. The resulting partition is derived offline and admits a generalization-bound interpretation.

Our contributions are summarized as follows: 
\begin{itemize}
    \item We propose a theoretical framework for phased-wise tuning of MLLMs in multi-framework WebUI code generation, where an offline method is used to obtain phase partition.
    \item We introduce a heterogeneity measure based on normalized codes and a compatibility measure based on offline pre-gradients, and use these two signals to automatically derive a phase partition and training schedule for multi-phase adapter finetuning.
    \item We build a multi-framework finetuning dataset based on DesignBench \cite{xiao2025designbench} and WebCode2M \cite{gui2025webcode2m}, and evaluate our method on DesignBench \cite{xiao2025designbench} and Design2Code \cite{si2025design2code}; the results consistently outperform baselines.
\end{itemize}

\section{Related works}
\subsection{WebUI Code Generation} 
Early front-end code generation approaches relied on end-to-end CNN–RNN image-to-sequence models that translated screenshots into intermediate framework-specific languages \cite{beltramelli2018pix2code, xu2021image2emmet}. Subsequent work moved toward direct HTML generation \cite{chen2018ui} and improved structural fidelity through explicit layout recognition \cite{cizotto2023web} or computer-vision pipelines such as OCR and object detection to recover hierarchical UI representations \cite{nguyen2015reverse, wu2021screen}. With the emergence of multimodal large language models, recent progress has been driven by large-scale synthetic training data \cite{laurenccon2024unlocking}, layout-aware hierarchical planning \cite{gui2025uicopilot, gui2025latcoder}, and decomposition-based generation strategies \cite{wan2025divide, wu2025mllm}, together with post-training and efficiency improvements \cite{xiao2025efficientuicoder, zhao2025vincicoder, chen2025breaking}. However, most existing studies focus on generation within a single frontend framework \cite{xiao2025designbench,zhou2025declarui,wu2026mulfcoder}. In practical deployment, applications often require supporting multiple frameworks that differ in syntax conventions, component abstractions, and engineering constraints. Directly transferring models trained for one framework to another—e.g., through prompt adaptation or lightweight fine-tuning—can lead to increased compilation errors and structural inconsistencies, highlighting the challenge of cross-framework robustness \cite{xiao2025designbench}.

\subsection{MLLM}
Multimodal large language models (MLLMs) extend pretrained language models to visual inputs by integrating a vision encoder with a language backbone through projection or connector modules that map visual features into token space \cite{alayrac2022flamingo, li2023blip}. Training typically proceeds via large-scale caption or interleaved image–text supervision, followed by visual instruction tuning to improve alignment and controllability \cite{liu2023visual, chen2024sharegpt4v}. Recent advances further enhance perception and reasoning through higher-resolution visual tokenization, extended context windows, and refined post-training strategies \cite{liu2024llavanext, team2024gemini, zhu2025internvl3}. 

\subsection{LLM Fine-Tuning}
Fine-tuning large language models has evolved from full-parameter updates to parameter-efficient strategies that adjust only a limited subset of parameters \cite{ding2023parameter,lialin2023scaling}. Methods such as adapters \cite{houlsby2019parameter}, prefix tuning \cite{li2021prefix}, and low-rank adaptation \cite{hu2022lora} introduce lightweight modules that enable task adaptation with reduced memory and computational overhead. These approaches have demonstrated strong performance across language and multimodal tasks and are widely used in practice \cite{xie2023difffit, sung2022vl}. Nevertheless, most fine-tuning methods are designed for single-task or single-framework scenarios \cite{houlsby2019parameter,li2021prefix,devlin2019bert}. When multiple frameworks or environments are optimized jointly, shared parameter updates may introduce gradient interference and negative transfer, particularly under heterogeneous output distributions or distinct engineering constraints \cite{liu2021conflict}. Addressing such cross-framework interaction requires training strategies that explicitly account for framework compatibility in addition to parameter efficiency \cite{standley2020tasks,kollias2021distribution, pfeiffer2021adapterfusion}.

\section{Method}

\subsection{Problem Setting}
Given a webpage screenshot $x$ and a target frontend framework identifier $f\in\{\texttt{React},\texttt{Vue},\texttt{Angular},\texttt{Vanilla}\}$, our goal is to generate runnable frontend code $y_f$ in the specified framework under its corresponding syntactic and compilation constraints, such that its rendered output matches the screenshot $x$ as closely as possible in both structural layout and visual appearance. 
We denote the training set by $\mathcal{D} = \{(x_i,y_i,f_i)\}_{i=1}^N$, where $N$ is the total number of training samples. 
For each framework $f$, we define the framework-specific subset as $\mathcal{D}_f = \{(x_i,y_i,f_i)\in\mathcal{D}\mid f_i=f\}$, which contains all pairs associated with framework $f$.

\subsection{Model Parameters and Loss}

We freeze the MLLM's parameters $\Theta$ and introduce adapters for $k$ frameworks with parameters $W = \{w^{s},w_1,\dots,w_k\}$, where $w^{s}$ denotes the shared parameters across all frameworks and $w_j$ denotes the framework-specific parameters. The modified MLLM is defined as:
\begin{equation}
\mathcal{F}(x;\Theta,W)=\mathrm{MLLM}(x;\Theta,{w^{s},w_1,\dots,w_k}).
\end{equation}  
To explicitly control the effective capacity introduced by adapters, we assume a bounded-complexity condition on $W$, namely that there exists a constant $\beta>0$ such that $||w||_2\le \beta$ for all $w\in W$. This constraint prevents any single framework-specific adapter from dominating the optimization dynamics, and it also enables a unified analysis in which the shared adapter $w^{s}$ and the framework-specific adapters $\{w_j\}_{j=1}^k$ are treated under the same norm budget.

Given framework-specific datasets, we finetune the adapters by minimizing the average expected risk across frameworks:
\begin{equation}
\mathcal{L}_{train}=\frac{1}{k}\sum_{j=1}^k \mathbb{E}_{(x,y,f)\sim \mathcal{D}j}\big[\ell(\mathcal{F}(x;\Theta,W),y)\big].
\end{equation}
Here, $\ell(\hat y,y)$ denotes a sequence-level supervised loss under teacher forcing. Although our evaluation focuses on compilation success, optimizing $\mathcal{L}_{train}$ provides a differentiable surrogate that encourages the generated code to follow the target framework syntax and conventions encoded in $\mathcal{D}_j$. Notably, the explicit framework index $j$ in the risk decomposition makes it possible to analyze how shared parameters $w^{s}$ and framework-specific parameters $w_j$ jointly affect the training dynamics across heterogeneous frameworks.

\begin{assumption}[Lipschitz]
The loss $\ell(\hat y,y)$ is $L$-Lipschitz with respect to $\hat y$. Moreover, there exists a constant $\gamma>0$ such that for any $W=(w^{s},{w_j})$ and $W^{'}=(w^{s'},{w'_j})$ and any input $x$,
\begin{equation}
\|\mathcal{F}(x;\Theta,W)-\mathcal{F}(x;\Theta,W^{'})\|
\le \gamma*\|W-W^{'}\|_{2,1} = \gamma(\|w^{s}-w^{s'}\|_2+\sum_{j=1}^k\|w_j-w_j^{'}\|_2)
\end{equation}
This assumption links perturbations of adapter parameters to changes in the model output, and it will be used to upper bound cross-framework deviations induced by different adapter updates.
\end{assumption}

\subsection{Heterogeneity}

For each framework dataset $j$, let $Q_j$ denote the empirical distribution induced by its normalized code representations. 
For two framework datasets $\mathcal{D}_i$ and $\mathcal{D}_j$, we define their framework heterogeneity as
\begin{equation}
\Delta_{i,j} = h(\mathcal{D}_i,\mathcal{D}_j)\triangleq \mathcal{H}(Q_i,Q_j),
\end{equation}
where $\mathcal{H}(\cdot,\cdot)\in[0,1]$ is a symmetric and bounded divergence between distributions. In this paper, we adopt the normalized Jensen-Shannon divergence. 
Here, $Q_j$ is not constructed from raw code strings directly, but from normalized code distributions, in order to reduce the influence of superficial stylistic variations on the resulting statistics. The details are deferred to Appendix~\ref{app:code_distribution}.
This definition measures the statistical shift between frameworks in the output space. If two frameworks admit more similar normalized code distributions, then their target code patterns, local structures, and compositional regularities are also more aligned, which makes stable transfer through shared parameters more likely.

\begin{assumption}[Risk Bridge for Proxy heterogeneity]
\label{ass:risk}
There exist a constant $\kappa\ge 0$ and a residual $\epsilon_h\ge 0$ such that for any admissible parameters and any pair of frameworks $\mathcal{D}_i$, $\mathcal{D}_j$, we have
\begin{equation}
\big|\mathcal{L}(\mathcal{D}_i;\Theta,W)-\mathcal{L}(\mathcal{D}_j;\Theta,W)\big|
\le
\kappa \Delta_{i,j}+\varepsilon_h.
\end{equation}

This assumption states that the heterogeneity defined on $Q(y)$ is not merely an empirical statistic, but can upper bound the risk shift between frameworks. 
The residual term $\varepsilon_h$ absorbs the portion of framework shift that cannot be fully explained by output distributions alone, such as input-side variation, latent semantic mismatch, or annotation noise.
\end{assumption}

\subsection{Offline Compatibility Estimation}
\label{sec:offline_compatibility}
We choose a backbone layer index $r$ and place the shared adapter parameter $w^s$ at the output of layer $r-1$, namely the same insertion point used during training. For each framework $\mathcal{D}_j$, the offline shared gradient at initialization is defined as
\begin{equation}
\bar g_j=\nabla_{w^s} \mathcal{L}(\mathcal{D}_j;\Theta,W=w^s,\{w\})\Big|_{w^{s}=0,w=0}.
\end{equation}
It represents the first-order update direction preferred by framework $\mathcal{D}_j$ on the shared adapter under MLLM and zero-initialized adapters.
The compatibility between two frameworks is defined by
\begin{equation}
S(\mathcal{D}_i,\mathcal{D}_j)
=
\frac{1+\cos(\bar g_i,\bar g_j)}{2}
\in [0,1].
\label{eq:offline_compatibility}
\end{equation}
A larger value indicates better alignment of the induced shared update directions, and thus stronger suitability for being grouped into the same phase.

For a candidate phase $P$, with weights $\{\alpha_j\}_{j\in P}$ satisfying $\alpha_j\ge 0$ and $\sum_{j\in P}\alpha_j=1$, we define the aggregate offline gradient and its compatibility margin as
\begin{equation}
\bar g_P=\sum_{j\in P}\alpha_j \bar g_j,
\qquad
\bar m(P)
=
\min_{j\in P}
\frac{\langle \bar g_j,\bar g_P\rangle}{\|\bar g_P\|_2^2}.
\label{eq:offline_margin}
\end{equation}
When $\bar m(P)>0$, the phase is first-order non-conflicting with respect to the shared adapter.
We only consider candidate phases satisfying $\|\bar g_P\|_2>0$; phases with vanishing aggregate offline gradients are treated as infeasible.

For the considered phase $P$, let $g_j$ denote the current shared gradient induced by framework $\mathcal{D}_j$, and let $g_P=\sum_{j\in P}\alpha_j g_j$ be its aggregate gradient.

\begin{assumption}[Offline-to-online directional stability]
\label{ass:compatibility_stability}
There exists $\varepsilon_g\ge 0$ such that $\|g_j-\bar g_j\|_2\le \varepsilon_g$ for all frameworks $\mathcal{D}_j$ in the current phase $P$.
\end{assumption}

\begin{lemma}[Compatibility-preserving shared update]
\label{lem:compatibility_preserving}
Assume that $\mathcal{L}(\mathcal{D}_j;\Theta,W)$ is $L_s$-smooth with respect to $w^s$ for every framework $\mathcal{D}_j$. Define the online compatibility margin by
\begin{equation}
m(P)
=
\min_{j\in P}
\frac{\langle g_j,g_P\rangle}{\|g_P\|_2^2}.
\label{eq:online_margin}
\end{equation}
If $m(P)>0$ and the shared update $w^{s,+}=w^s-\eta g_P$ uses a step size $\eta\le 2m(P)/L_s$, then
\begin{equation}
\mathcal{L}(\mathcal{D}_j;\Theta,W^{+})
\le
\mathcal{L}(\mathcal{D}_j;\Theta,W),
\qquad
\forall j\in P.
\label{eq:no_harm_phase}
\end{equation}
\end{lemma}

Lemma~\ref{lem:compatibility_preserving} shows that compatibility is not merely a heuristic similarity score. Once the online margin is positive, the aggregated shared update is descent-compatible for all frameworks in the phase. The justification for using the offline gradients $\{\bar g_j\}$ as surrogates of their online counterparts is deferred to Appendix~\ref{app:compatibility_proofs}.

\subsection{Phase-wise Generalization Bounds}

\begin{theorem}[Phase-wise concurrent generalization bound]
\label{thm:phase_generalization}
Let $P\subseteq \{1,\dots,k\}$ be a candidate phase and let $n_P=\sum_{j\in P}|\mathcal{D}_j|$. Under the Lipschitz assumption, the bounded-complexity condition on $W$, and Assumption~\ref{ass:risk}, for any $\delta>0$, with probability at least $1-\delta$, every admissible model satisfies
\begin{equation}
\sum_{j\in P}
\alpha_j \mathcal{L}(\mathcal{D}_j;\Theta,W)
\le
\sum_{j\in P}
\alpha_j \widehat{\mathcal{L}}(\mathcal{D}_j;\Theta,W)
+
\Gamma(P)
+
\kappa
\sum_{\substack{i<j\\ i,j\in P}}
h(\mathcal{D}_i,\mathcal{D}_j)
+
\epsilon_h
+
O\left(
\sqrt{\frac{\ln(1/\delta)}{n_P}}
\right).
\label{eq:phase_bound}
\end{equation}
where $\widehat{\mathcal{L}}$ is empirical risk and 
\begin{equation}
\Gamma(P)
\triangleq
2L\gamma
\left(
\beta+\sum_{j\in P}\alpha_j \beta
\right)
\le 4L\gamma\beta.
\label{eq:phase_gamma}
\end{equation}
\end{theorem}

Theorem~\ref{thm:phase_generalization} shows that the phase risk is controlled by the empirical loss, the adapter complexity, and the total internal heterogeneity inside the phase. 
Let $\Pi=(P_1,\dots,P_M)$ and $\bigcup_{t=1}^M P_t=\{1,\dots,k\}.$
For each phase $P_t$, let $\{\alpha_j^t\}_{j\in P_t}$ be nonnegative weights such that $\sum_{j\in P_t}\alpha_j^t=1$.

The heterogeneity term in Theorem~\ref{thm:phase_generalization} motivates penalizing internally divergent phases. In contrast, the compatibility term is introduced through Lemma~\ref{lem:compatibility_preserving}, which shows that positively aligned shared gradients yield descent-compatible shared-adapter updates for all frameworks in a phase. Thus, our partition objective combines the heterogeneity penalty from the generalization bound with the compatibility reward from the shared-update analysis.

Motivated by this bound and the offline compatibility score in Eq.~\eqref{eq:offline_compatibility}, we define the partition objective
\begin{equation}
G(P_1,\dots,P_M)
=
-\sum_{t=1}^M
\left[
\sum_{\substack{i,j\in P_t\\ i<j}} h(\mathcal{D}_i,\mathcal{D}_j)
-
\lambda
\sum_{\substack{i,j\in P_t\\ i<j}} S(\mathcal{D}_i,\mathcal{D}_j)
+
\mu_s\|w^{s,t}\|_2^2
+
\mu_f\sum_{j\in P_t}\|w_j^t\|_2^2
\right],
\label{eq:partition_objective}
\end{equation}
where $\lambda,\mu_s,\mu_f>0$ and $M$ is the number of phase. A larger value of $G$ therefore corresponds to smaller internal heterogeneity, larger constructive compatibility, and lower per-phase adapter complexity.
The norm-based capacity term in Eq.(\ref{eq:partition_objective}) serves as a computable surrogate for the complexity contribution $\Gamma(P_t)$ in Eq.(\ref{eq:phase_bound}), since the bounded adapter complexity condition controls both quantities through the shared and framework-specific adapter norm budgets.

\begin{theorem}[Multi-Phase Partition with Compatibility-Capacity Maximization]
\label{thm:partition_selection}

Let
\begin{equation}
(P_1^\star,\ldots,P_M^\star)
=
\arg\max_{\Pi:\ \bar m(P_t)>C_{P_t}\epsilon_g,\ \forall t}
G(P_1,\ldots,P_M).
\end{equation}
$C_{P_t}$ is the stability constant from Proposition \ref{prop:margin_transfer} that converts offline margin into online margin.
Suppose that every selected phase $P_t^\star$ satisfies the condition of Lemma~\ref{lem:compatibility_preserving}. Then, under the assumptions of Theorem~\ref{thm:phase_generalization}, the phase-wise training trajectory satisfies
\begin{equation}
R_{\max}(P_1^\star,\dots,P_M^\star)
\le
\big[1-G(P_1^\star,\dots,P_M^\star)\big]_+
+
O\!\left(\sqrt{\frac{\ln(1/\delta)}{N}}\right),
\label{eq:partition_final_bound}
\end{equation}
where $R_{\max}\triangleq \max_{t\in\{1,\ldots,M\}}
\sum_{j\in P_t^\star}\alpha_j^t L(D_j;\Theta,W^t)$,
$N=\sum_{j=1}^{k}|D_j|$, $G$ denotes the monotone normalized objective, and $[u]_+=\max\{0,u\}$.
\end{theorem}

Theorem~\ref{thm:partition_selection} directly explains the training rule of our method. We first choose the partition that maximizes Eq.~\eqref{eq:partition_objective}, and then perform phase-wise adapter optimization on the selected phases.
The gating rule is defined in Appendix~\ref{app:gate}.
The process of algorithm is shown in Appendix~\ref{app:alg}.

\newpage

\input{exp}

\section{Conclusion}

We presented a phase-wise adapter tuning method for multi-framework WebUI code generation, using offline heterogeneity from normalized code distributions and compatibility from shared-adapter pre-gradient alignment to guide framework partitioning. The resulting training schedule, together with lightweight per-framework gating, reduces cross-framework interference without requiring additional online gradient computation.

\bibliographystyle{IEEEtran}
\bibliography{reference}

\newpage
\appendix

\section{Algorithm process}
\label{app:alg}

\begin{algorithm}[h!]
\caption{Phase-wise adapter tuning}
\label{alg:phase_tuning}
\begin{algorithmic}[1]
\Require Frozen MLLM parameters $\Theta$; framework datasets $\{\mathcal{D}_j\}_{j=1}^{k}$; shared adapter $w^s$; framework-specific adapters $\{w_j\}_{j=1}^{k}$; number of phases $M$.
\State Compute normalized code distributions and heterogeneity scores $\Delta_{i,j}$ for all framework pairs.
\State Compute offline pre-gradients $\{\bar g_j\}_{j=1}^{k}$ and compatibility scores $S(\mathcal{D}_i,\mathcal{D}_j)$.
\State Construct the feasible partition set using $\bar m(P_t)>C_{P_t}\epsilon_g$.
\State Select $\Pi^\star=(P_1^\star,\ldots,P_M^\star)$ by maximizing $G$ in Eq.~(13).
\State Compute framework gates $\{a_m\}_{m=1}^{k}$ from average heterogeneity and compatibility.
\For{$t=1$ to $M$}
    \State Sample training batches only from frameworks in $P_t^\star$.
    \State Update the shared adapter $w^s$ using the gate-weighted shared update.
    \State Update framework-specific adapters $\{w_j:j\in P_t^\star\}$.
    \State Keep framework-specific adapters $\{w_j:j\notin P_t^\star\}$ frozen.
\EndFor
\State \Return tuned adapters $w^s,\{w_j\}_{j=1}^{k}$.
\end{algorithmic}
\end{algorithm}

\section{Construction of Normalized Code Distributions}
\label{app:code_distribution}

To estimate output-space differences across domains in a stable and reproducible manner, we do not construct empirical distributions directly from raw code strings. Instead, we first normalize the code and then estimate its $n$-gram statistics. For a target code sample $y$ from domain $\mathcal{D}_j$, let its normalized form be denoted by
\begin{equation}
\widetilde{y}=\mathrm{Norm}(y).
\end{equation}

The normalization mapping $\mathrm{Norm}(\cdot)$ consists of four steps. First, we standardize whitespace, indentation, line breaks, and quotation styles, and remove comments or formatting artifacts that do not affect functionality. This reduces the sensitivity of distribution estimation to code style. Second, we abstract away local variable names, temporary function names, auto-generated class names, and random identifiers, so that free naming choices do not dominate the resulting statistics. Third, we perform lightweight normalization for local patterns that are semantically equivalent but syntactically heterogeneous across domains, which improves comparability while preserving the essential code structure. Finally, we tokenize the normalized code and estimate its empirical unigram and bigram distributions.

Formally, let $Q_j^{(1)}$ and $Q_j^{(2)}$ denote the unigram and bigram empirical distributions of domain $\mathcal{D}_j$ lifted to a shared disjoint support, respectively. We define empirical code distribution of domain $j$ as
\begin{equation}
Q_j=(1-\omega)Q_j^{(1)}+\omega Q_j^{(2)},
\qquad \omega\in[0,1].
\end{equation}
By default, we set $\omega=\tfrac{1}{2}$.

Given two domains $\mathcal{D}_i$ and $\mathcal{D}_j$, their discrepancy is computed by the normalized Jensen--Shannon divergence
\begin{equation}
\Delta_{i,j}
=
\frac{1}{\log 2}
\left[
\frac{1}{2}\mathrm{KL}(Q_i\|M_{ij})
+
\frac{1}{2}\mathrm{KL}(Q_j\|M_{ij})
\right],
\qquad
M_{ij}=\frac{1}{2}(Q_i+Q_j),
\end{equation}
which ensures that $\Delta_{i,j}\in[0,1]$.

The purpose of this construction is not to remove all output differences across domains, but to suppress superficial perturbations that are irrelevant to the task while preserving the statistics that reflect target code patterns, structural motifs, and local compositional regularities. The resulting $Q_j$ is therefore both computable and stable, making it a suitable proxy for the discrepancy term in stage partition.

\section{Proofs and Justifications for Offline Compatibility}
\label{app:compatibility_proofs}

\subsection{Proof of Lemma~\ref{lem:compatibility_preserving}}

For a fixed phase $P$, recall the shared update
\begin{equation}
w^{s,+}=w^{s}-\eta g_P,
\qquad
g_P=\sum_{j\in P}\alpha_j g_j.
\end{equation}
Fix an arbitrary framework $\mathcal{D}_j\in P$ and define
\begin{equation}
F_j(w^s)\triangleq \mathcal{L}(\mathcal{D}_j;\Theta,W),
\end{equation}
where all non-shared parameters are fixed at their current values. Since $F_j$ is $L_s$-smooth with respect to $w^s$, we have
\begin{equation}
F_j(w^{s,+})
\le
F_j(w^{s})
+
\langle \nabla_{w^s}F_j(w^{s}),\, w^{s,+}-w^{s}\rangle
+
\frac{L_s}{2}\|w^{s,+}-w^{s}\|_2^2.
\label{eq:smoothness_proof_1}
\end{equation}
Substituting
\begin{equation}
w^{s,+}-w^{s}=-\eta g_P
\end{equation}
and using
\begin{equation}
\nabla_{w^s}F_j(w^{s})=g_j,
\end{equation}
we obtain
\begin{equation}
F_j(w^{s,+})
\le
F_j(w^{s})
-
\eta\langle g_j,g_P\rangle
+
\frac{L_s\eta^2}{2}\|g_P\|_2^2.
\label{eq:smoothness_proof_2}
\end{equation}
By the definition of $m(P)$,
\begin{equation}
\langle g_j,g_P\rangle
\ge
m(P)\|g_P\|_2^2,
\qquad \forall j\in P.
\end{equation}
Therefore,
\begin{equation}
F_j(w^{s,+})
\le
F_j(w^{s})
-
\eta m(P)\|g_P\|_2^2
+
\frac{L_s\eta^2}{2}\|g_P\|_2^2.
\end{equation}
Rearranging yields
\begin{equation}
F_j(w^{s,+})-F_j(w^{s})
\le
\left(
-\eta m(P)+\frac{L_s\eta^2}{2}
\right)\|g_P\|_2^2.
\end{equation}
Hence, whenever
\begin{equation}
\eta\le \frac{2\,m(P)}{L_s},
\end{equation}
the right-hand side is non-positive, which implies
\begin{equation}
\mathcal{L}(\mathcal{D}_j;\Theta,W^{+})
\le
\mathcal{L}(\mathcal{D}_j;\Theta,W),
\qquad
\forall j\in P.
\end{equation}
This proves Lemma~\ref{lem:compatibility_preserving}.

\subsection{Why can offline gradients serve as surrogates of online shared gradients?}

The key observation is that the offline gradient is computed at exactly the same insertion location as the shared adapter used in training, namely the output of layer $r-1$. Even before the shared adapter is explicitly optimized, the loss still admits a well-defined first-order sensitivity with respect to a virtual perturbation at that location.

\begin{proposition}[Virtual perturbation equivalence]
\label{prop:virtual_perturbation}
Let $H^{r-1}(x)$ denote the hidden representation at the output of layer $r-1$. Introduce a virtual perturbation variable $\Delta$ at the shared insertion point and consider the perturbed representation
\begin{equation}
H^{r-1}(x)\mapsto H^{r-1}(x)+\Delta.
\end{equation}
Then, at $\Delta=0$, the gradient of the loss with respect to the perturbation equals the gradient with respect to a zero-initialized additive shared adapter:
\begin{equation}
\nabla_{\Delta}\mathcal{L}(\mathcal{D}_j;\Theta,\Delta)\Big|_{\Delta=0}
=
\nabla_{w^s}\mathcal{L}(\mathcal{D}_j;\Theta,W)\Big|_{w^s=0,\;w=0}.
\label{eq:virtual_equivalence}
\end{equation}
\end{proposition}

\begin{proof}
Consider the additive shared adapter
\begin{equation}
A_{w^s}(H^{r-1})=H^{r-1}+w^s.
\end{equation}
At initialization $w^s=0$, the adapter output coincides with the unmodified representation, and the Jacobian of $A_{w^s}$ with respect to $w^s$ is the identity map. By the chain rule,
\begin{equation}
\nabla_{w^s}\mathcal{L}
=
\left(
\frac{\partial A_{w^s}(H^{r-1})}{\partial w^s}
\right)^\top
\nabla_{A_{w^s}(H^{r-1})}\mathcal{L}.
\end{equation}
Evaluating at $w^s=0$ gives
\begin{equation}
\nabla_{w^s}\mathcal{L}\Big|_{w^s=0}
=
\nabla_{H^{r-1}}\mathcal{L},
\end{equation}
which is exactly the gradient with respect to the virtual perturbation $\Delta$ at $\Delta=0$.
\end{proof}

Proposition~\ref{prop:virtual_perturbation} explains why the offline shared gradient remains well-defined even before the shared adapter is explicitly trained.

\begin{proposition}[Sample-average characterization]
\label{prop:sample_average_gradient}
Assume that differentiation and expectation can be interchanged. Then the framework-level offline shared gradient satisfies
\begin{equation}
\bar g_j
=
\nabla_{w^s}\mathbb{E}_{(x,y)\sim\mathcal{D}_j}
\big[\ell(x,y;\Theta,W)\big]\Big|_{w^s=0,\;w=0}
=
\mathbb{E}_{(x,y)\sim\mathcal{D}_j}
\left[
\nabla_{w^s}\ell(x,y;\Theta,W)\Big|_{w^s=0,\;w=0}
\right].
\label{eq:sample_average_gradient}
\end{equation}
\end{proposition}

\begin{proof}
The result follows directly from the interchange of differentiation and expectation applied to the empirical loss over $\mathcal{D}_j$.
\end{proof}

Proposition~\ref{prop:sample_average_gradient} justifies estimating $\bar g_j$ by averaging sample-wise gradients within each framework.

\begin{proposition}[First-order consistency for general shared adapters]
\label{prop:first_order_consistency}
Suppose the shared adapter inserted at layer $r-1$ admits the local expansion
\begin{equation}
A_{w^s}(H^{r-1})
=
H^{r-1}
+
J(H^{r-1})\,w^s
+
\mathcal{O}(\|w^s\|_2^2)
\end{equation}
around $w^s=0$, where $J(H^{r-1})$ is the Jacobian of the adapter output with respect to $w^s$. Then
\begin{equation}
\nabla_{w^s}\mathcal{L}(\mathcal{D}_j;\Theta,W)\Big|_{w^s=0,\;w=0}
=
J(H^{r-1})^\top \nabla_{H^{r-1}}\mathcal{L}(\mathcal{D}_j;\Theta).
\label{eq:first_order_projection}
\end{equation}
Hence the hidden-state sensitivity at the insertion point determines the first-order shared update direction up to a fixed linear projection.
\end{proposition}

\begin{proof}
By the chain rule,
\begin{equation}
\nabla_{w^s}\mathcal{L}
=
\left(
\frac{\partial A_{w^s}(H^{r-1})}{\partial w^s}
\right)^\top
\nabla_{A_{w^s}(H^{r-1})}\mathcal{L}.
\end{equation}
Evaluating this identity at $w^s=0$ and using the stated local expansion yields \eqref{eq:first_order_projection}.
\end{proof}

Proposition~\ref{prop:first_order_consistency} shows that the offline gradient proxy is not restricted to a purely additive adapter parameterization. It remains valid for any shared adapter that is linearizable around the initialization point.

\begin{proposition}[Gradient drift bound under bounded adapter updates]
\label{prop:gradient_drift}
Assume that, for each framework $\mathcal{D}_j$, the shared gradient $\nabla_{w^s}\mathcal{L}(\mathcal{D}_j;\Theta,W)$ is $L_g$-Lipschitz with respect to $W$. If the adapter trajectory stays in a bounded neighborhood,
\begin{equation}
\|W\|_2\le \rho,
\end{equation}
then for every framework $j$,
\begin{equation}
\|g_j-\bar g_j\|_2\le L_g \rho.
\label{eq:gradient_drift}
\end{equation}
\end{proposition}

\begin{proof}
By the $L_g$-Lipschitz continuity of the shared gradient,
\begin{equation}
\left\|
\nabla_{w^s}\mathcal{L}(\mathcal{D}_j;\Theta,W)
-
\nabla_{w^s}\mathcal{L}(\mathcal{D}_j;\Theta,0)
\right\|_2
\le
L_g\|W-0\|_2.
\end{equation}
Since
\begin{equation}
\bar g_j
=
\nabla_{w^s}\mathcal{L}(\mathcal{D}_j;\Theta,W)\Big|_{w^s=0,\;w=0},
\end{equation}
the claim follows from $\|W\|_2\le \rho$.
\end{proof}

Proposition~\ref{prop:gradient_drift} explains why the offline gradient can serve as a meaningful surrogate of the online shared gradient in the frozen-backbone regime. Since only adapters are updated and their trajectory is norm-controlled, the online shared gradients remain close to their initialization counterparts.

\begin{proposition}[From offline margin to online margin]
\label{prop:margin_transfer}
For a phase $P$, define
\begin{equation}
\bar g_P=\sum_{j\in P}\alpha_j \bar g_j,
\qquad
\bar m(P)
=
\min_{j\in P}
\frac{\langle \bar g_j,\bar g_P\rangle}{\|\bar g_P\|_2^2}.
\end{equation}
Assume that $\|g_j-\bar g_j\|_2\le \varepsilon_g$ for all $j\in P$, and that there exist constants
\begin{equation}
\|\bar g_j\|_2\le G_P,
\qquad
\|\bar g_P\|_2\ge \gamma_P>\varepsilon_g.
\end{equation}
Then there exists a phase-dependent constant $C_P>0$ such that
\begin{equation}
m(P)\ge \bar m(P)-C_P\varepsilon_g.
\label{eq:margin_transfer}
\end{equation}
In particular, if $\bar m(P)>C_P\varepsilon_g$, then $m(P)>0$.
\end{proposition}

\begin{proof}
Let $\Delta_j=g_j-\bar g_j$ and $\Delta_P=g_P-\bar g_P=\sum_{j\in P}\alpha_j\Delta_j$. Then $\|\Delta_P\|_2\le \varepsilon_g$. For any $j\in P$,
\begin{equation}
\langle g_j,g_P\rangle
=
\langle \bar g_j,\bar g_P\rangle
+
\langle \Delta_j,\bar g_P\rangle
+
\langle \bar g_j,\Delta_P\rangle
+
\langle \Delta_j,\Delta_P\rangle.
\end{equation}
Therefore,
\begin{equation}
\left|
\langle g_j,g_P\rangle
-
\langle \bar g_j,\bar g_P\rangle
\right|
\le
\varepsilon_g(\gamma_P+G_P)+\varepsilon_g^2.
\end{equation}
Moreover,
\begin{equation}
\|g_P\|_2
\ge
\|\bar g_P\|_2-\|\Delta_P\|_2
\ge
\gamma_P-\varepsilon_g,
\end{equation}
and thus
\begin{equation}
\|g_P\|_2^2\ge (\gamma_P-\varepsilon_g)^2.
\end{equation}
Combining the perturbation bound for the numerator with the lower bound for the denominator yields
\begin{equation}
\frac{\langle g_j,g_P\rangle}{\|g_P\|_2^2}
\ge
\frac{\langle \bar g_j,\bar g_P\rangle}{\|\bar g_P\|_2^2}
-
C_P\varepsilon_g
\end{equation}
for some constant $C_P$ depending only on $(G_P,\gamma_P)$. Taking the minimum over $j\in P$ proves \eqref{eq:margin_transfer}.
\end{proof}

This justifies using \eqref{eq:offline_compatibility} and \eqref{eq:offline_margin} during phase construction before the actual multi-phase optimization begins.

\section{Proofs of the Generalization Results}
\label{app:generalization_proofs}

\subsection{Proof of Theorem 3.1}

\begin{proof}
Fix a candidate phase $P\subseteq\{1,\ldots,k\}$ and define the phase-level weighted risk
\begin{equation} 
\mathcal{L}_P(\Theta,W)
=
\sum_{j\in P}\alpha_j \mathcal{L}(\mathcal{D}_j;\Theta,W),
\qquad
\widehat {\mathcal{L}}_P(\Theta,W)
=
\sum_{j\in P}\alpha_j \widehat {\mathcal{L}}(\mathcal{D}_j;\Theta,W).
\end{equation}
By Assumption 3.1 and the bounded adapter complexity condition, the admissible adapter class has bounded effective capacity. Standard uniform-convergence arguments for Lipschitz losses over norm-restricted parameter classes imply that, with probability at least $1-\delta$,
\begin{equation}
\mathcal{L}_P(\Theta,W)
\le
\widehat {\mathcal{L}}_P(\Theta,W)
+
\Gamma(P)
+
O\left(
\sqrt{\frac{\ln(1/\delta)}{n_P}}
\right),
\end{equation}
where
\begin{equation}
\Gamma(P)
=
2L\gamma
\left(
\beta+\sum_{j\in P}\alpha_j\beta
\right).
\end{equation}
Next, by the risk bridge assumption for proxy heterogeneity, the residual shift among frameworks inside the same phase is controlled by their pairwise heterogeneity. Absorbing constant normalizers into $\kappa$, we have
\begin{equation}
\sum_{j\in P}\alpha_j \mathcal{L}(\mathcal{D}_j;\Theta,W)
\le
\sum_{j\in P}\alpha_j \widehat {\mathcal{L}}(\mathcal{D}_j;\Theta,W)
+
\Gamma(P)
+
\kappa
\sum_{\substack{i<j\\i,j\in P}}
h(\mathcal{D}_i,\mathcal{D}_j)
+
\epsilon_h
+
O\left(
\sqrt{\frac{\ln(1/\delta)}{n_P}}
\right).
\end{equation}
This is exactly the claimed phase-wise concurrent generalization bound.
\end{proof}

\subsection{Proof of Theorem 3.2}

\begin{proof}
Let $\Pi=(P_1,\ldots,P_M)$ be any feasible phase partition satisfying
\begin{equation}
\bar m(P_t)>C_{P_t}\epsilon_g,\qquad \forall t.
\end{equation}
By Proposition \ref{prop:margin_transfer}, this condition implies $m(P_t)>0$ for every phase. Lemma 3.1 then guarantees that the aggregate shared-adapter update in phase $P_t$ is descent-compatible for all frameworks in that phase.

For a fixed phase $P_t$, applying Theorem 3.1 gives, with high probability,
\begin{equation}
\sum_{j\in P_t}
\alpha_j^t
\mathcal{L}(\mathcal{D}_j;\Theta,W^t)
\le
\sum_{j\in P_t}
\alpha_j^t
\widehat {\mathcal{L}}(\mathcal{D}_j;\Theta,W^t)
+
\Gamma(P_t)
+
\kappa
\sum_{\substack{i<j\\i,j\in P_t}}
h(\mathcal{D}_i,\mathcal{D}_j)
+
\epsilon_h
+
O\left(
\sqrt{\frac{\ln(1/\delta)}{n_t}}
\right).
\end{equation}
The norm-based capacity term $\mathrm{Cap}(P_t)$ upper bounds the complexity contribution and the phase-wise adapter budget. Moreover, by the compatibility calibration condition, the descent benefit associated with compatible shared updates is lower bounded by
\begin{equation}
\lambda
\sum_{\substack{i<j\\i,j\in P_t}}
S(\mathcal{D}_i,\mathcal{D}_j).
\end{equation}
Therefore the phase risk is upper bounded by
\begin{equation}
\sum_{j\in P_t}
\alpha_j^t
\mathcal{L}(\mathcal{D}_j;\Theta,W^t)
\le
C_t
+
\sum_{\substack{i<j\\i,j\in P_t}}
h(\mathcal{D}_i,\mathcal{D}_j)
-
\lambda
\sum_{\substack{i<j\\i,j\in P_t}}
S(\mathcal{D}_i,\mathcal{D}_j)
+
\mathrm{Cap}(P_t)
+
O\left(
\sqrt{\frac{\ln(1/\delta)}{n_t}}
\right),
\end{equation}
where $C_t$ absorbs the empirical loss bound, $\Gamma(P_t)$, and constants independent of the partition, $\mathrm{Cap}(P_t) = \mu_s\|w^{s,t}\|_2^2
+
\mu_f\sum_{j\in P_t}\|w_j^t\|_2^2$.

Taking the maximum over phases and using $\max_t a_t\le \sum_t a_t$, we obtain
\begin{equation}
R_{\max}(\Pi)
\le
1
-
G(\Pi)
+
O\left(
\sqrt{\frac{\ln(1/\delta)}{N}}
\right),
\end{equation}
where $N=\sum_{j=1}^{k}|\mathcal{D}_j|$.

Finally, since $\widetilde G$ is a monotone normalization of $G$ over the feasible partition set, maximizing $G$ and maximizing $\widetilde G$ yield the same partition. Hence the optimizer
\begin{equation}
\Pi^\star
=
\arg\max_{\Pi:\bar m(P_t)>C_{P_t}\epsilon_g,\forall t}
G(\Pi)
\end{equation}
minimizes the partition-dependent part of the upper bound. This gives
\begin{equation}
R_{\max}(\Pi^\star)
\le
\big[
1- \widetilde G(\Pi^\star)
\big]_+
+
O\left(
\sqrt{\frac{\ln(1/\delta)}{N}}
\right),
\end{equation}
which proves the theorem.
\end{proof}

\section{The details of 
Offline relationship-guided gating.}

\label{app:gate}
In addition to determining the phase partition, the offline heterogeneity and compatibility scores are also used to assign a framework-level shared-update gate. For each framework $m$, we compute its average heterogeneity and compatibility with the other frameworks:
\begin{equation}
\bar{\Delta}_m
=
\frac{1}{k-1}
\sum_{i\ne m}
h(\mathcal{D}_i,D_m),
\qquad
\bar{S}_m
=
\frac{1}{k-1}
\sum_{i\ne m}
S(\mathcal{D}_i,D_m).
\end{equation}
The gate value $a_m\in[0,1]$ is obtained from these two offline statistics through a fixed calibration rule:
\begin{equation}
a_m
=
\mathrm{Calib}(\bar{S}_m,\bar{\Delta}_m),
\end{equation}
where $\mathrm{Calib}(\cdot,\cdot)$ is monotonic with respect to the two signals, increasing with $\bar{S}_m$ and decreasing with $\bar{\Delta}_m$. Thus, frameworks that are more compatible with the others and less heterogeneous receive larger shared-update weights, while more isolated frameworks rely more strongly on their framework-specific adapters.
In our implementation, $\mathrm{Calib}$ is instantiated by first min-max normalizing $\bar S_m$ and $\bar\Delta_m$ across frameworks and then linearly combining them with equal weights, $a_m = \tfrac{1}{2}\,\hat S_m + \tfrac{1}{2}\,(1-\hat\Delta_m)$, followed by clipping to $[0,1]$, where $\hat S_m$ and $\hat\Delta_m$ denote the normalized signals.
Because Angular is simultaneously the most heterogeneous ($\hat\Delta_m = 1$) and the least compatible ($\hat S_m = 0$) framework, this rule assigns it $a_m = 0$: its shared updates are fully disabled and it relies entirely on its framework-specific adapter.

During training, the framework identifier only performs a lookup over the precomputed gate table. For a training sample $(x_i,y_i,f_i)$, let $m=\iota(f_i)$ be its framework index. The framework-specific adapter $w_m$ is selected by this index, and the same index retrieves the scalar gate $a_m$ for the shared adapter update:
\begin{equation}
w^s
\leftarrow
w^s
-
\eta\,a_m
\nabla_{w^s}
\ell(\mathcal{F}(x_i;\Theta,W),y_i),
\end{equation}
while the framework-specific adapter is updated normally:
\begin{equation}
w_m
\leftarrow
w_m
-
\eta
\nabla_{w_m}
\ell(\mathcal{F}(x_i;\Theta,W),y_i).
\end{equation}
This gate is computed offline and is not an input-dependent learned routing module.

\section{Additional Ablation Studies}
\label{app:additional_ablations}

\paragraph{Gradient Layer Position.}
Table~\ref{tab:ablation_layer} reports the effect of choosing different layers for offline synergy gradient computation.
Our default layer~15 achieves the best result ($0.7950 \pm 0.0027$) with the lowest synergy variance.
Earlier layers (8--12) show higher variance ($0.0063$--$0.0082$), as low-level visual features carry little framework-specific signal, whereas layers within the Adapter insertion range (18--22) perform worse (CLIP $\leq 0.7823$) and also exhibit elevated estimate variance ($0.0048$--$0.0062$), suggesting that compatibility estimated from deep, decoding-stage-dominated representations is less stable; layer~15 balances visual-semantic abstraction and task specificity.

\begin{table}[h]
\centering
\caption{Effect of gradient computation layer (mean$\pm$std over 3 seeds).}
\label{tab:ablation_layer}
\begin{tabular}{lccc}
\toprule
\textbf{Layer} & \textbf{CLIP (Avg.)} & \textbf{CSR (Avg.)} & \textbf{Synergy Std.} \\
\midrule
Layer 8                  & $0.7763{\pm}0.0079$ & $0.8714{\pm}0.0061$ & $0.0082$ \\
Layer 10                 & $0.7798{\pm}0.0065$ & $0.8741{\pm}0.0051$ & $0.0071$ \\
Layer 12                 & $0.7878{\pm}0.0049$ & $0.8802{\pm}0.0043$ & $0.0063$ \\
Layer 14                 & $0.7924{\pm}0.0033$ & $0.8917{\pm}0.0030$ & $0.0041$ \\
\midrule
\textbf{Layer 15 (ours)} & $\mathbf{0.7950{\pm}0.0027}$ & $\mathbf{0.9007{\pm}0.0024}$ & $0.0031$ \\
\midrule
Layer 18                 & $0.7823{\pm}0.0038$ & $0.8771{\pm}0.0035$ & $0.0048$ \\
Layer 20                 & $0.7774{\pm}0.0044$ & $0.8728{\pm}0.0040$ & $0.0053$ \\
Layer 22                 & $0.7736{\pm}0.0056$ & $0.8698{\pm}0.0051$ & $0.0062$ \\
\bottomrule
\end{tabular}
\end{table}

\paragraph{Discrepancy Distance Function.}
Table~\ref{tab:ablation_dist} compares three choices of distance function $d(\cdot,\cdot)$ for computing $\Delta$.
JS Divergence (our default) achieves the best mean ($0.7950 \pm 0.0027$).
Sinkhorn-Wasserstein yields a comparable mean ($0.7931 \pm 0.0030$) but incurs $3.7\times$ the offline computation time.
MMD shows slightly lower performance ($0.7912 \pm 0.0033$), attributed to sensitivity to the bandwidth hyperparameter on finite code token vocabularies.

\begin{table}[h]
\centering
\caption{Discrepancy distance function ablation (mean$\pm$std over 3 seeds).}
\label{tab:ablation_dist}
\begin{tabular}{lccc}
\toprule
\textbf{Distance Function} & \textbf{CLIP (Avg.)} & \textbf{CSR (Avg.)} & \textbf{Offline Comp.} \\
\midrule
MMD                                    & $0.7912{\pm}0.0033$ & $0.8901{\pm}0.0029$ & $1.4\times$ \\
Sinkhorn-Wasserstein                   & $0.7931{\pm}0.0030$ & $0.8921{\pm}0.0028$ & $3.7\times$ \\
\midrule
\textbf{Ours (JS Divergence)}          & $\mathbf{0.7950{\pm}0.0027}$ & $\mathbf{0.9007{\pm}0.0024}$ & $1.0\times$ \\
\bottomrule
\end{tabular}
\end{table}

\paragraph{Code Normalization Strength.}
Table~\ref{tab:ablation_norm} ablates the strength of the normalization operator $\mathcal{N}_f$.
Weak normalization (whitespace and formatting only) leads to noisier discrepancy estimates, degrading CLIP by $0.014$ points.
Strong normalization (additionally removing framework-specific syntax tokens) over-smooths the distribution, degrading CLIP by $0.018$ points.
Our default moderate normalization achieves the best balance.

\begin{table}[h]
\centering
\caption{Code normalization strength ablation.}
\label{tab:ablation_norm}
\begin{tabular}{lcc}
\toprule
\textbf{Normalization} & \textbf{CLIP (Avg.)} & \textbf{CSR (Avg.)} \\
\midrule
Weak (format only)                     & 0.7812 & 0.8747 \\
Strong (remove framework syntax)       & 0.7769 & 0.8694 \\
\midrule
\textbf{Ours (moderate)}               & \textbf{0.7950} & \textbf{0.9007} \\
\bottomrule
\end{tabular}
\end{table}

\section{Editing and Repair Task Results}
\label{app:editing_repair}

This appendix reports results on the DesignBench Editing and Repair tasks.
Both tasks use the MLLM-as-Judge scoring protocol from DesignBench~\cite{xiao2025designbench}, where a separate frontier MLLM (GPT-4o) evaluates the output on a 0--10 scale covering visual fidelity, structural correctness, and framework-specific syntax adherence.

\paragraph{Editing Task.}
The Editing task provides the model with an existing code snippet and a screenshot of the desired modification target.
The model must produce code that matches the visual target while preserving the structural parts of the original code that are not subject to change.
Table~\ref{tab:editing_results} reports MLLM-as-Judge scores for each framework.

\begin{table}[h]
\centering
\caption{MLLM-as-Judge scores on the DesignBench Editing task (0--10 scale; mean$\pm$std over 3 seeds for fine-tuned methods).}
\label{tab:editing_results}
\resizebox{\linewidth}{!}{%
\begin{tabular}{lccccc}
\toprule
\textbf{Method} & \textbf{Vanilla} & \textbf{React} & \textbf{Vue} & \textbf{Angular} & \textbf{Avg.} \\
\midrule
Claude-3.7           & $7.42$ & $7.83$ & $7.61$ & $5.94$ & $7.20$ \\
GPT-4o               & $7.23$ & $7.41$ & $7.39$ & $5.61$ & $6.91$ \\
Qwen2.5-VL-7B        & $6.81$ & $5.12$ & $5.34$ & $1.03$ & $4.58$ \\
\midrule
ScreenCoder          & $7.34{\pm}0.08$ & $7.52{\pm}0.11$ & $7.48{\pm}0.09$ & $6.12{\pm}0.14$ & $7.12{\pm}0.10$ \\
EfficientUICoder     & $7.12{\pm}0.09$ & $7.23{\pm}0.13$ & $7.19{\pm}0.11$ & $5.87{\pm}0.16$ & $6.85{\pm}0.12$ \\
\midrule
\textbf{Ours} & $\mathbf{7.68{\pm}0.07}$ & $\mathbf{7.94{\pm}0.09}$ & $\mathbf{7.87{\pm}0.08}$ & $\mathbf{6.89{\pm}0.12}$ & $\mathbf{7.60{\pm}0.09}$ \\
\bottomrule
\end{tabular}%
}
\end{table}

Our method achieves the best macro-average score ($7.60$), outperforming ScreenCoder by $+0.48$ points and the zero-shot Claude-3.7 baseline by $+0.40$ points.
The most pronounced improvement is on Angular ($+0.77$ over ScreenCoder), where the framework-specific Adapter captures Angular's reactive binding syntax necessary for structurally faithful code edits.
The gap between our method and Claude-3.7 is narrower on Editing than on Generation, likely because editing tasks require preserving existing code structure---a capability in which large frontier models retain an advantage through in-context pattern matching.

\paragraph{Repair Task.}
The Repair task provides a code snippet containing compilation errors and requires the model to produce a corrected, compilable version.
Table~\ref{tab:repair_results} reports MLLM-as-Judge scores.

\begin{table}[h]
\centering
\caption{MLLM-as-Judge scores on the DesignBench Repair task (0--10 scale; mean$\pm$std over 3 seeds for fine-tuned methods).}
\label{tab:repair_results}
\resizebox{\linewidth}{!}{%
\begin{tabular}{lccccc}
\toprule
\textbf{Method} & \textbf{Vanilla} & \textbf{React} & \textbf{Vue} & \textbf{Angular} & \textbf{Avg.} \\
\midrule
Claude-3.7           & $7.91$ & $8.12$ & $7.98$ & $7.03$ & $7.76$ \\
GPT-4o               & $7.78$ & $7.93$ & $7.82$ & $6.87$ & $7.60$ \\
Qwen2.5-VL-7B        & $7.34$ & $6.23$ & $6.47$ & $2.14$ & $5.55$ \\
\midrule
ScreenCoder          & $7.62{\pm}0.08$ & $7.81{\pm}0.10$ & $7.74{\pm}0.09$ & $6.58{\pm}0.13$ & $7.44{\pm}0.10$ \\
EfficientUICoder     & $7.43{\pm}0.10$ & $7.61{\pm}0.12$ & $7.52{\pm}0.11$ & $6.34{\pm}0.15$ & $7.23{\pm}0.12$ \\
\midrule
\textbf{Ours} & $\mathbf{8.03{\pm}0.06}$ & $\mathbf{8.24{\pm}0.08}$ & $\mathbf{8.17{\pm}0.07}$ & $\mathbf{7.52{\pm}0.11}$ & $\mathbf{7.99{\pm}0.08}$ \\
\bottomrule
\end{tabular}%
}
\end{table}

On the Repair task, our method achieves macro-average $7.99$, which surpasses Claude-3.7 ($7.76$) by $+0.23$ points.
The improvement is strongest on Angular ($+0.94$ over ScreenCoder), where compilation errors frequently involve template binding syntax errors that the framework-specific Adapter learns to correct reliably.
The Repair task rewards compilation-aware generation---a property consistent with our compilation-oriented evaluation goal, where the gating mechanism reduces shared Adapter updates for frameworks with high discrepancy, preventing cross-framework syntax confusion that commonly leads to compilation failures.

\section{Discussions}
\label{app:disc}

\paragraph{Limitations.}
The proposed phase partition and gating strategy rely on offline heterogeneity and compatibility estimates, so their reliability may decrease when the training data are sparse, highly imbalanced, or substantially different from the target deployment distribution. 

\paragraph{Broader Impacts.}
This work can reduce the engineering effort required to translate UI designs into executable frontend code, especially when developers need to support multiple frontend frameworks. It may also make rapid prototyping and framework migration more accessible to small teams and non-expert users.
\newpage
\input{checklist.tex}

\end{document}

%% file: exp.tex
\section{Experiments}
\label{sec:experiments}

\subsection{Experimental Setup}
\label{sec:exp_setup}

\paragraph{Datasets and Metrics.}
We evaluate on two public benchmarks.
\textbf{DesignBench}~\cite{xiao2025designbench} supports four front-end frameworks (Vanilla, React, Vue, Angular) and covers Generation, Editing, and Repair tasks; we report Generation results in the main paper, evaluated on the held-out test split summarized in Table~\ref{tab:dataset_stats}, and Editing/Repair in Appendix~\ref{app:editing_repair}.
\textbf{Design2Code}~\cite{si2025design2code} provides 484 real-world Vanilla webpage samples with five fine-grained automatic metrics.
For training, we combine screenshot--code pairs from DesignBench~\cite{xiao2025designbench} and WebCode2M~\cite{gui2025webcode2m} into a multi-framework fine-tuning set (Table~\ref{tab:dataset_stats}).
On DesignBench we report \textbf{CLIP} visual similarity and \textbf{CSR} (Compilation Success Rate); on Design2Code we report Block Match, Text, Position, Color, and CLIP.
CLIP is computed between the rendered page of each generated code and the target screenshot; since uncompilable code admits no faithful rendering, CLIP is averaged over samples whose code compiles successfully, whereas CSR is computed over all test samples of the framework.

\begin{table}[t]
\centering
\caption{Dataset statistics per framework.}
\label{tab:dataset_stats}
\begin{tabular}{lcccc}
\toprule
\textbf{Framework} & \textbf{Train} & \textbf{Val} & \textbf{Test} & \textbf{Avg.\ Tokens} \\
\midrule
Vanilla  & 2,814 & 312 & 300 & 1,023 \\
React    & 2,773 & 308 & 295 & 1,187 \\
Vue      & 2,681 & 298 & 284 & 1,142 \\
Angular  & 2,594 & 288 & 277 & 1,309 \\
\midrule
Total    & 10,862 & 1,206 & 1,156 & 1,165 \\
\bottomrule
\end{tabular}
\end{table}

\paragraph{Model and Training.}
We use \texttt{Qwen2.5-VL-7B-Instruct} as the frozen backbone.
One shared Adapter ($r=16$, $\alpha=32$) is inserted at layers 16--27; four framework-specific Adapters ($r=8$, $\alpha=16$) are inserted at layers 20--27.
The offline partition algorithm yields \textbf{Phase~1}: $\{$React, Vue$\}$ ($S=0.79$, $\Delta=0.18$); \textbf{Phase~2}: $\{$Vanilla, Angular$\}$ ($S=0.31$, $\Delta=0.73$).
We train 3 epochs per phase with batch size 32, learning rate $2{\times}10^{-4}$, and cosine warmup.
Offline compatibility (synergy) gradients are computed at layer 15 over $K=100$ batches per framework.
All fine-tuning experiments use 3 random seeds; we report mean scores in the main comparison and ablation tables, and mean$\pm$std in the additional ablations (Appendix~\ref{app:additional_ablations}) and the MLLM-as-judge evaluations (Appendix~\ref{app:editing_repair}).

\paragraph{Baselines.}
We compare against: (1)~\textit{frontier MLLMs (zero-shot)}: Claude-3.7~\cite{anthropic2025claude37}, GPT-4o~\cite{openai2024gpt4o}, Gemini-2.0-Flash~\cite{google2025gemini20}, Llama-3.2-11B~\cite{meta2024llama32}, Qwen2.5-VL-7B~\cite{wang2024qwen25vl}; (2)~\textit{structured generation}: ScreenCoder~\cite{wu2025mllm}; (3)~\textit{efficiency-oriented}: EfficientUICoder~\cite{xiao2025efficientuicoder}; (4)~\textit{specialized baselines on Design2Code}: GPT-4V and the Design2Code-18B model~\cite{si2025design2code}; and (5)~\textit{internal ablations} (Section~\ref{sec:ablation}).

\subsection{Main Results}
\label{sec:main_results}

\paragraph{Multi-framework Generation on DesignBench.}
Table~\ref{tab:main_designbench} shows that our method improves both visual fidelity and compilation reliability in the multi-framework generation setting. The gains are especially meaningful because they are obtained across frameworks with substantially different syntax systems, rather than being concentrated in a single easy framework. Compared with strong zero-shot MLLMs, our method closes the gap between general visual-code understanding and framework-aware executable generation. The improvement on Angular is particularly informative: this framework imposes stricter component and template constraints, and the result suggests that framework-specific Adapters together with gated shared updates help the model learn compilation-critical conventions. Compared with the ungated shared-specific baseline, the full method further demonstrates that offline relationship signals contribute beyond adding more trainable parameters.

\begin{table}[t]
\centering
\caption{Results on DesignBench Generation task. Best results are \textbf{bold}; second best are \underline{underlined}.}
\label{tab:main_designbench}
\resizebox{\linewidth}{!}{%
\begin{tabular}{lcccccccccc}
\toprule
\multirow{2}{*}{\textbf{Method}} & \multicolumn{2}{c}{\textbf{Vanilla}} & \multicolumn{2}{c}{\textbf{React}} & \multicolumn{2}{c}{\textbf{Vue}} & \multicolumn{2}{c}{\textbf{Angular}} & \multicolumn{2}{c}{\textbf{Avg.}} \\
\cmidrule(lr){2-3}\cmidrule(lr){4-5}\cmidrule(lr){6-7}\cmidrule(lr){8-9}\cmidrule(lr){10-11}
 & CLIP & CSR & CLIP & CSR & CLIP & CSR & CLIP & CSR & CLIP & CSR \\
\midrule
Claude-3.7         & \underline{0.7832} & \textbf{1.0000} & \underline{0.7891} & \underline{0.9492} & \underline{0.7934} & 0.6655 & 0.6223 & 0.6029 & \underline{0.7470} & 0.8044 \\
GPT-4o             & 0.7635 & \textbf{1.0000} & 0.6763 & 0.9119 & 0.7734 & 0.6092 & 0.5867 & 0.5379 & 0.7000 & 0.7648 \\
Gemini-2.0         & 0.7412 & \textbf{1.0000} & 0.7611 & 0.8407 & 0.6997 & 0.1408 & 0.6006 & 0.4838 & 0.7007 & 0.6163 \\
Llama-3.2-11B      & 0.6304 & \textbf{1.0000} & 0.6401 & 0.4915 & 0.3241 & 0.5810 & 0.5834 & 0.5415 & 0.5445 & 0.6535 \\
Qwen2.5-VL-7B      & 0.7411 & \textbf{1.0000} & 0.6875 & 0.1085 & 0.6452 & 0.1197 & 0.5186 & 0.0866 & 0.6481 & 0.3287 \\
\midrule
ScreenCoder        & 0.7821 & \textbf{1.0000} & 0.7534 & 0.9232 & 0.7609 & \underline{0.7887} & 0.6234 & \underline{0.7822} & 0.7300 & \underline{0.8735} \\
EfficientUICoder   & 0.7623 & \textbf{1.0000} & 0.7312 & 0.8915 & 0.7423 & 0.7629 & 0.5934 & 0.7232 & 0.7073 & 0.8444 \\
\midrule
Specific Adapter only      & 0.7612 & \textbf{1.0000} & 0.7389 & 0.8621 & 0.7521 & 0.7617 & 0.6814 & 0.7196 & 0.7334 & 0.8359 \\
Shared+Specific (no gate)  & 0.7701 & \textbf{1.0000} & 0.7523 & 0.8791 & 0.7634 & 0.7735 & \underline{0.7012} & 0.7389 & 0.7468 & 0.8479 \\
\midrule
\textbf{Ours}              & \textbf{0.7956} & \textbf{1.0000} & \textbf{0.8124} & \textbf{0.9582} & \textbf{0.8231} & \textbf{0.8310} & \textbf{0.7489} & \textbf{0.8135} & \textbf{0.7950} & \textbf{0.9007} \\
\bottomrule
\end{tabular}%
}
\end{table}

\paragraph{Validation on Design2Code.}
Table~\ref{tab:main_design2code} evaluates whether the proposed multi-framework tuning procedure remains effective on a Vanilla-only benchmark. The results show that the method does not sacrifice standard HTML/CSS generation ability while learning multiple frontend frameworks. Instead, the shared training phases appear to improve layout and visual reconstruction, suggesting that cross-framework data provide useful structural supervision even when the evaluation target is Vanilla code. The only metric where a frontier model remains slightly ahead is text matching, which is less directly tied to framework syntax and more sensitive to language-level copying. Overall, the Design2Code results indicate that our framework-level specialization does not come at the cost of general webpage generation quality.

\begin{table}[t]
\centering
\caption{Results on Design2Code benchmark. Best results are \textbf{bold}; second best are \underline{underlined}.}
\label{tab:main_design2code}
\begin{tabular}{lccccc}
\toprule
\textbf{Method} & \textbf{Block Match} & \textbf{Text} & \textbf{Position} & \textbf{Color} & \textbf{CLIP} \\
\midrule
GPT-4o              & \underline{93.0} & \textbf{98.2} & \underline{85.5} & \underline{84.1} & \underline{90.4} \\
Claude-3.7          & 90.2             & 97.5          & 77.9             & 71.4             & 87.0 \\
GPT-4V              & 85.8             & 97.4          & 80.5             & 73.3             & 86.9 \\
Design2Code-18B     & 78.5             & 96.4          & 74.3             & 67.0             & 85.8 \\
ScreenCoder         & 84.3             & 97.1          & 79.2             & 75.6             & 87.3 \\
EfficientUICoder    & 81.7             & 96.8          & 77.4             & 72.3             & 86.1 \\
\midrule
\textbf{Ours}       & \textbf{94.7}    & \underline{97.9} & \textbf{87.2} & \textbf{85.6}    & \textbf{91.3} \\
\bottomrule
\end{tabular}
\end{table}

\subsection{Ablation Study}
\label{sec:ablation}

\paragraph{Architecture Ablation.}
Table~\ref{tab:ablation_arch} separates the effect of shared knowledge, framework-specific capacity, phase-wise scheduling, and gating. The specific-only variant lacks a mechanism for sharing reusable layout and component patterns, while the shared-only variant cannot sufficiently preserve framework-dependent syntax conventions. Combining shared and specific Adapters improves this balance, but the ungated version still treats all frameworks as equally suitable for shared updates. Adding phase-wise training already improves the optimization path by controlling which frameworks are learned together, and the full gated version further refines shared updates according to offline relationship scores. 

\begin{table}[htbp]
\centering
\begin{minipage}{0.48\linewidth}
\centering
\caption{Architecture ablation.}
\label{tab:ablation_arch}
\resizebox{\linewidth}{!}{%
\begin{tabular}{lcc}
\toprule
\textbf{Variant} & \textbf{CLIP} & \textbf{CSR} \\
\midrule
Specific Adapter only                    & 0.7334 & 0.8359 \\
Shared Adapter only                      & 0.7408 & 0.8412 \\
Shared+Specific (no gate, no phase)      & 0.7468 & 0.8479 \\
Shared+Specific (no gate, with phase)    & 0.7741 & 0.8712 \\
\midrule
\textbf{Ours (gate + phase)}             & \textbf{0.7950} & \textbf{0.9007} \\
\bottomrule
\end{tabular}%
}
\end{minipage}\hfill
\begin{minipage}{0.48\linewidth}
\centering
\caption{Gating component ablation.}
\label{tab:ablation_gate}
\resizebox{\linewidth}{!}{%
\begin{tabular}{lcc}
\toprule
\textbf{Gating Variant} & \textbf{CLIP} & \textbf{CSR} \\
\midrule
Discrepancy-only gate                  & 0.7803 & 0.8762 \\
Synergy-only gate                      & 0.7779 & 0.8734 \\
Uniform gate ($a_m = 0.5$)             & 0.7698 & 0.8681 \\
Random gate                            & 0.7643 & 0.8627 \\
\midrule
\textbf{Ours (combined gate)}          & \textbf{0.7950} & \textbf{0.9007} \\
\bottomrule
\end{tabular}%
}
\end{minipage}
\end{table}

\paragraph{Gating Component Ablation.}
Table~\ref{tab:ablation_gate} examines whether the gate should be driven by heterogeneity, compatibility, or both. The discrepancy-only and synergy-only variants each improve over unstructured gating, showing that both output-distribution distance and gradient-direction alignment contain useful training signals. However, neither signal alone is sufficient: discrepancy identifies when shared updates may be risky, while synergy identifies when shared updates are likely to be beneficial. The combined gate performs best because it jointly suppresses updates for highly heterogeneous frameworks and preserves shared learning for compatible ones. The weaker performance of uniform and random gates---both of which fall below the ungated phase-wise baseline (CLIP $0.7741$ / CSR $0.8712$ in Table~\ref{tab:ablation_arch})---further shows that the effect is not merely due to adding a scalar weight: mis-calibrated gating is harmful rather than neutral, so the gain depends on the framework-specific values inferred from offline statistics. All gating variants in Table~\ref{tab:ablation_gate} use the same two-phase schedule as our full method.

\paragraph{Phase Partition Ablation.}
Table~\ref{tab:ablation_phase} evaluates whether the partition itself matters, beyond the presence of multiple training phases. The single-phase baseline mixes all frameworks and therefore exposes the shared Adapter to conflicting output patterns throughout training. Random or manually altered two-phase partitions perform worse than the automatically selected schedule, indicating that grouping decisions cannot be reduced to arbitrary batching. 
Increasing to $M=3$ remains competitive but does not surpass the selected two-phase schedule, while fully separating all frameworks weakens useful cross-framework transfer.

\begin{table}[htbp]
\centering
\begin{minipage}{0.48\linewidth}
\centering
\caption{Phase partition ablation.}
\label{tab:ablation_phase}
\resizebox{\linewidth}{!}{%
\begin{tabular}{lcc}
\toprule
\textbf{Phase Configuration} & \textbf{CLIP} & \textbf{CSR} \\
\midrule
$M=1$                                & 0.7608 & 0.8649 \\
$M=2$: reversed order                & 0.7817 & 0.8789 \\
$M=2$: Vanilla+React / Vue+Angular   & 0.7834 & 0.8812 \\
$M=3$                                & 0.7921 & 0.8951 \\
$M=4$                                & 0.7801 & 0.8709 \\
\midrule
\textbf{Ours ($M=2$)}                & \textbf{0.7950} & \textbf{0.9007} \\
\bottomrule
\end{tabular}%
}
\end{minipage}\hfill
\begin{minipage}{0.48\linewidth}
\centering
\caption{Synergy estimation strategy ablation.}
\label{tab:ablation_synergy}
\resizebox{\linewidth}{!}{%
\begin{tabular}{lccc}
\toprule
\textbf{Synergy Variant} & \textbf{CLIP} & \textbf{CSR} & \textbf{Extra Cost} \\
\midrule
Online gradient synergy                & 0.7916 & 0.8908 & $+23\%$ \\
No synergy ($S = \mathbf{0}$)         & 0.7782 & 0.8745 & $0$ \\
\midrule
\textbf{Ours (offline pre-gradient)}   & \textbf{0.7950} & \textbf{0.9007} & $+2\%$ \\
\bottomrule
\end{tabular}%
}
\end{minipage}
\end{table}

\paragraph{Synergy Estimation: Offline vs.\ Online.}
Table~\ref{tab:ablation_synergy} compares the proposed offline pre-gradient estimate with online gradient statistics. The offline estimate achieves comparable performance while avoiding the substantial overhead of repeatedly collecting online gradient information during training. 
This supports the central assumption behind our compatibility measure.
The no-synergy variant performs worse, confirming that phase construction needs a positive compatibility signal rather than relying only on distributional heterogeneity. Thus, offline pre-gradients provide a cost-effective proxy for deciding which frameworks should share training phases.
Additional ablations on gradient layer selection, discrepancy distance function, and code normalization strength are reported in Appendix~\ref{app:additional_ablations}.

\subsection{Analysis}
\label{sec:analysis}

\begin{figure}[htbp]
\centering
\begin{minipage}{0.52\linewidth}
\centering
\includegraphics[width=\linewidth]{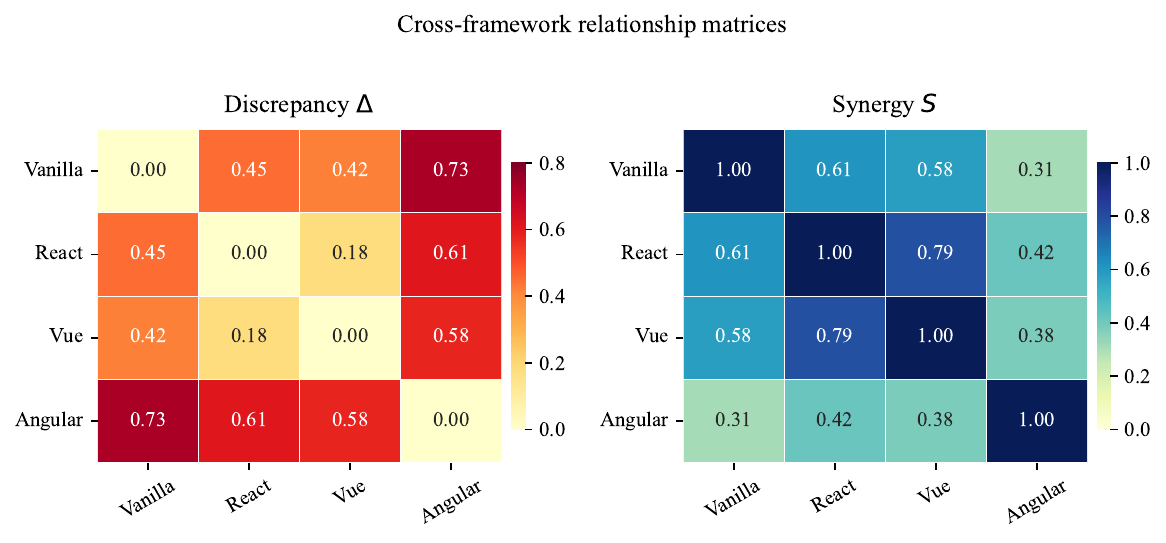}
\caption{Offline discrepancy ($\Delta$, i.e., heterogeneity, left) and synergy ($S$, i.e., compatibility, right) matrices.}
\label{fig:heatmap}
\end{minipage}\hfill
\begin{minipage}{0.45\linewidth}
\centering
\includegraphics[width=\linewidth]{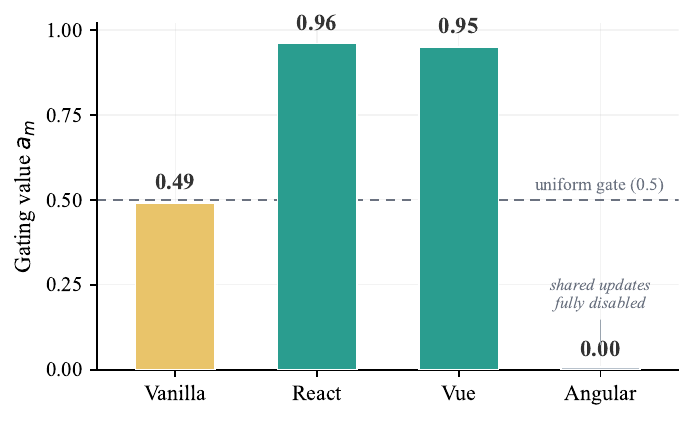}
\caption{Per-framework gating values $a_m$ derived deterministically from the offline statistics; the dashed line marks the uniform gate ($a_m{=}0.5$).}
\label{fig:gate_hist}
\end{minipage}
\vspace{-1em}
\end{figure}

\paragraph{Discrepancy--Synergy Statistics.}
Figure~\ref{fig:heatmap} visualizes the two offline relationship matrices used by the phase selection algorithm. The matrices reveal that heterogeneity and compatibility are not redundant: 
React and Vue form the clearest compatible group, which explains why they are placed together in the first phase. 
Vanilla and Angular show a much larger discrepancy and weaker synergy; 
their grouping is therefore better interpreted as the residual second phase under a two-phase budget, with subsequent gating disabling shared updates for the most isolated framework.

\paragraph{Gating Distribution and Theoretical Correlation.}
Figure~\ref{fig:gate_hist} shows that the framework-level gates reflect the offline relationship scores: React and Vue, which are mutually compatible and close in normalized code distributions, receive the largest gates ($0.96$ and $0.95$), Vanilla an intermediate gate ($0.49$), and Angular a fully closed gate ($0.00$), as it is simultaneously the most heterogeneous and the least compatible framework, so its training relies entirely on the framework-specific Adapter. 
The observed gate pattern qualitatively aligns with the framework-wise gains over the no-gate baseline, providing empirical support for the gating mechanism.

\subsection{Error Analysis}
\label{sec:error_analysis}

\begin{figure}[htbp]
\centering
\includegraphics[width=0.75\linewidth]{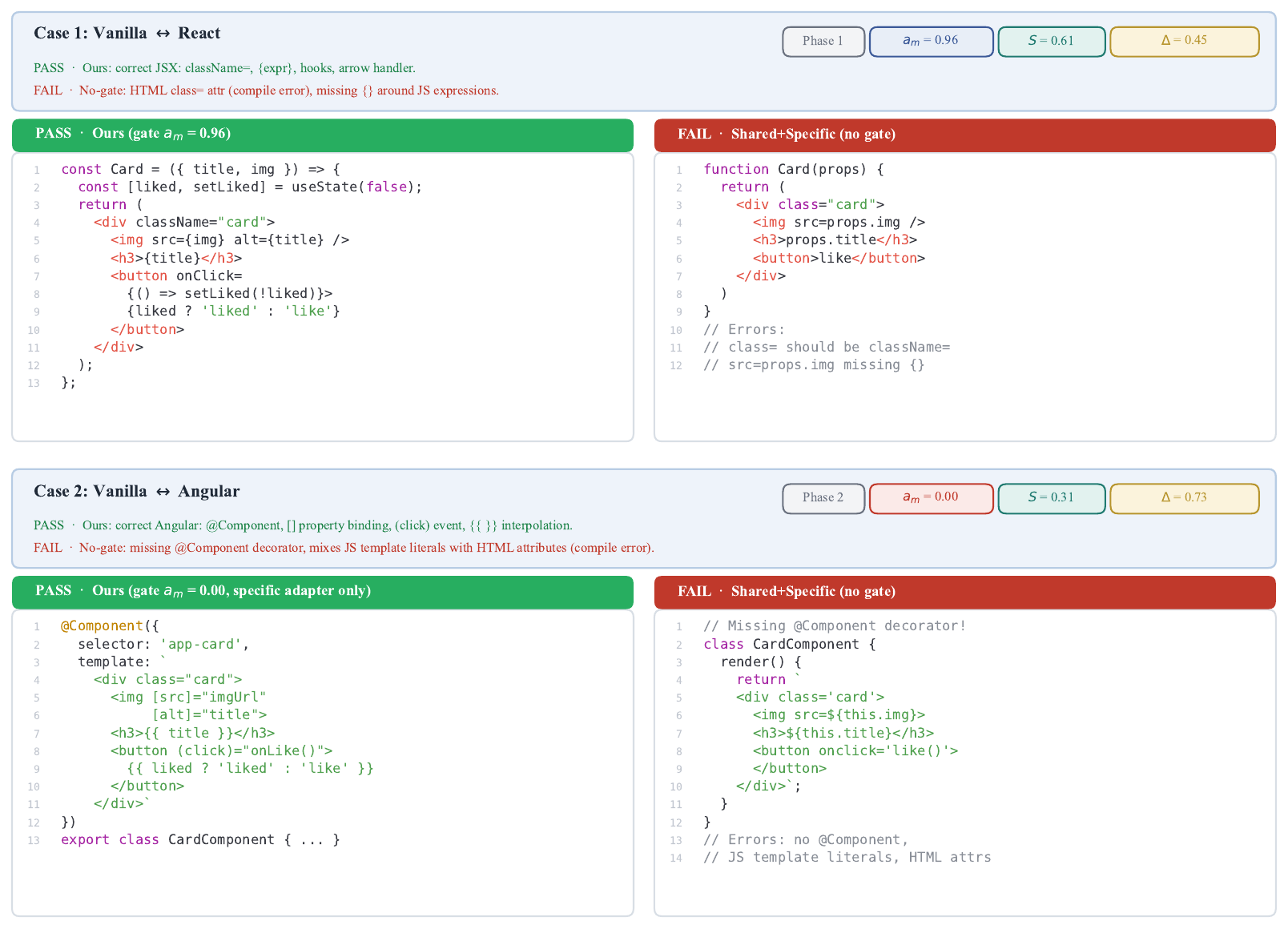}
\vspace{-1em}
\caption{Qualitative code generation examples stratified by ($\Delta$, $S$). Each row shows our method (center, green border) vs.\ Shared+Specific no-gate baseline (right, red border).}
\label{fig:qualitative}
\end{figure}

\paragraph{Qualitative Examples.}
Figure~\ref{fig:qualitative} presents representative examples stratified by ($\Delta$, $S$) values.
The top row (Vanilla vs.\ React, $\Delta=0.45$, $S=0.61$) shows our method correctly generating JSX functional components with hooks while preserving shared structural layout.
The bottom row (Vanilla vs.\ Angular, $\Delta=0.73$, $S=0.31$) shows the gating mechanism fully disabling shared updates ($a_m=0.00$) so the specific Adapter correctly captures Angular template binding syntax.

%% file: checklist.tex
\section*{NeurIPS Paper Checklist}

\begin{enumerate}

\item {\bf Claims}
    \item[] Question: Do the main claims made in the abstract and introduction accurately reflect the paper's contributions and scope?
    \item[] Answer: \answerYes{} %
    \item[] Justification: The abstract and introduction sections reflect the paper’s contributions and
scope in the paper.
    \item[] Guidelines:
    \begin{itemize}
        \item The answer \answerNA{} means that the abstract and introduction do not include the claims made in the paper.
        \item The abstract and/or introduction should clearly state the claims made, including the contributions made in the paper and important assumptions and limitations. A \answerNo{} or \answerNA{} answer to this question will not be perceived well by the reviewers. 
        \item The claims made should match theoretical and experimental results, and reflect how much the results can be expected to generalize to other settings. 
        \item It is fine to include aspirational goals as motivation as long as it is clear that these goals are not attained by the paper. 
    \end{itemize}

\item {\bf Limitations}
    \item[] Question: Does the paper discuss the limitations of the work performed by the authors?
    \item[] Answer: \answerYes{} %
    \item[] Justification: Limitations are discussed in \ref{app:disc}.
    \item[] Guidelines:
    \begin{itemize}
        \item The answer \answerNA{} means that the paper has no limitation while the answer \answerNo{} means that the paper has limitations, but those are not discussed in the paper. 
        \item The authors are encouraged to create a separate ``Limitations'' section in their paper.
        \item The paper should point out any strong assumptions and how robust the results are to violations of these assumptions (e.g., independence assumptions, noiseless settings, model well-specification, asymptotic approximations only holding locally). The authors should reflect on how these assumptions might be violated in practice and what the implications would be.
        \item The authors should reflect on the scope of the claims made, e.g., if the approach was only tested on a few datasets or with a few runs. In general, empirical results often depend on implicit assumptions, which should be articulated.
        \item The authors should reflect on the factors that influence the performance of the approach. For example, a facial recognition algorithm may perform poorly when image resolution is low or images are taken in low lighting. Or a speech-to-text system might not be used reliably to provide closed captions for online lectures because it fails to handle technical jargon.
        \item The authors should discuss the computational efficiency of the proposed algorithms and how they scale with dataset size.
        \item If applicable, the authors should discuss possible limitations of their approach to address problems of privacy and fairness.
        \item While the authors might fear that complete honesty about limitations might be used by reviewers as grounds for rejection, a worse outcome might be that reviewers discover limitations that aren't acknowledged in the paper. The authors should use their best judgment and recognize that individual actions in favor of transparency play an important role in developing norms that preserve the integrity of the community. Reviewers will be specifically instructed to not penalize honesty concerning limitations.
    \end{itemize}

\item {\bf Theory assumptions and proofs}
    \item[] Question: For each theoretical result, does the paper provide the full set of assumptions and a complete (and correct) proof?
    \item[] Answer: \answerYes{} %
    \item[] Justification: Full assumptions and proofs are provided in Appendix~\ref{app:compatibility_proofs} and Appendix~\ref{app:generalization_proofs}.
    \item[] Guidelines:
    \begin{itemize}
        \item The answer \answerNA{} means that the paper does not include theoretical results. 
        \item All the theorems, formulas, and proofs in the paper should be numbered and cross-referenced.
        \item All assumptions should be clearly stated or referenced in the statement of any theorems.
        \item The proofs can either appear in the main paper or the supplemental material, but if they appear in the supplemental material, the authors are encouraged to provide a short proof sketch to provide intuition. 
        \item Inversely, any informal proof provided in the core of the paper should be complemented by formal proofs provided in appendix or supplemental material.
        \item Theorems and Lemmas that the proof relies upon should be properly referenced. 
    \end{itemize}

    \item {\bf Experimental result reproducibility}
    \item[] Question: Does the paper fully disclose all the information needed to reproduce the main experimental results of the paper to the extent that it affects the main claims and/or conclusions of the paper (regardless of whether the code and data are provided or not)?
    \item[] Answer: \answerYes{} %
    \item[] Justification: Implementation details are given in Section~\ref{sec:exp_setup}, Appendix~\ref{app:additional_ablations}, and Algorithm~\ref{alg:phase_tuning}.
    \item[] Guidelines:
    \begin{itemize}
        \item The answer \answerNA{} means that the paper does not include experiments.
        \item If the paper includes experiments, a \answerNo{} answer to this question will not be perceived well by the reviewers: Making the paper reproducible is important, regardless of whether the code and data are provided or not.
        \item If the contribution is a dataset and\slash or model, the authors should describe the steps taken to make their results reproducible or verifiable. 
        \item Depending on the contribution, reproducibility can be accomplished in various ways. For example, if the contribution is a novel architecture, describing the architecture fully might suffice, or if the contribution is a specific model and empirical evaluation, it may be necessary to either make it possible for others to replicate the model with the same dataset, or provide access to the model. In general. releasing code and data is often one good way to accomplish this, but reproducibility can also be provided via detailed instructions for how to replicate the results, access to a hosted model (e.g., in the case of a large language model), releasing of a model checkpoint, or other means that are appropriate to the research performed.
        \item While NeurIPS does not require releasing code, the conference does require all submissions to provide some reasonable avenue for reproducibility, which may depend on the nature of the contribution. For example
        \begin{enumerate}
            \item If the contribution is primarily a new algorithm, the paper should make it clear how to reproduce that algorithm.
            \item If the contribution is primarily a new model architecture, the paper should describe the architecture clearly and fully.
            \item If the contribution is a new model (e.g., a large language model), then there should either be a way to access this model for reproducing the results or a way to reproduce the model (e.g., with an open-source dataset or instructions for how to construct the dataset).
            \item We recognize that reproducibility may be tricky in some cases, in which case authors are welcome to describe the particular way they provide for reproducibility. In the case of closed-source models, it may be that access to the model is limited in some way (e.g., to registered users), but it should be possible for other researchers to have some path to reproducing or verifying the results.
        \end{enumerate}
    \end{itemize}

\item {\bf Open access to data and code}
    \item[] Question: Does the paper provide open access to the data and code, with sufficient instructions to faithfully reproduce the main experimental results, as described in supplemental material?
    \item[] Answer: \answerYes{} %
    \item[] Justification: We will release our code after this paper is accepted.
    \item[] Guidelines:
    \begin{itemize}
        \item The answer \answerNA{} means that paper does not include experiments requiring code.
        \item Please see the NeurIPS code and data submission guidelines (\url{https://neurips.cc/public/guides/CodeSubmissionPolicy}) for more details.
        \item While we encourage the release of code and data, we understand that this might not be possible, so \answerNo{} is an acceptable answer. Papers cannot be rejected simply for not including code, unless this is central to the contribution (e.g., for a new open-source benchmark).
        \item The instructions should contain the exact command and environment needed to run to reproduce the results. See the NeurIPS code and data submission guidelines (\url{https://neurips.cc/public/guides/CodeSubmissionPolicy}) for more details.
        \item The authors should provide instructions on data access and preparation, including how to access the raw data, preprocessed data, intermediate data, and generated data, etc.
        \item The authors should provide scripts to reproduce all experimental results for the new proposed method and baselines. If only a subset of experiments are reproducible, they should state which ones are omitted from the script and why.
        \item At submission time, to preserve anonymity, the authors should release anonymized versions (if applicable).
        \item Providing as much information as possible in supplemental material (appended to the paper) is recommended, but including URLs to data and code is permitted.
    \end{itemize}

\item {\bf Experimental setting/details}
    \item[] Question: Does the paper specify all the training and test details (e.g., data splits, hyperparameters, how they were chosen, type of optimizer) necessary to understand the results?
    \item[] Answer: \answerYes{} %
    \item[] Justification: See Section~\ref{sec:exp_setup} and Appendix~\ref{app:additional_ablations}.
    \item[] Guidelines:
    \begin{itemize}
        \item The answer \answerNA{} means that the paper does not include experiments.
        \item The experimental setting should be presented in the core of the paper to a level of detail that is necessary to appreciate the results and make sense of them.
        \item The full details can be provided either with the code, in appendix, or as supplemental material.
    \end{itemize}

\item {\bf Experiment statistical significance}
    \item[] Question: Does the paper report error bars suitably and correctly defined or other appropriate information about the statistical significance of the experiments?
    \item[] Answer: \answerYes{} %
    \item[] Justification: The reported results have run three times.
    \item[] Guidelines:
    \begin{itemize}
        \item The answer \answerNA{} means that the paper does not include experiments.
        \item The authors should answer \answerYes{} if the results are accompanied by error bars, confidence intervals, or statistical significance tests, at least for the experiments that support the main claims of the paper.
        \item The factors of variability that the error bars are capturing should be clearly stated (for example, train/test split, initialization, random drawing of some parameter, or overall run with given experimental conditions).
        \item The method for calculating the error bars should be explained (closed form formula, call to a library function, bootstrap, etc.)
        \item The assumptions made should be given (e.g., Normally distributed errors).
        \item It should be clear whether the error bar is the standard deviation or the standard error of the mean.
        \item It is OK to report 1-sigma error bars, but one should state it. The authors should preferably report a 2-sigma error bar than state that they have a 96\% CI, if the hypothesis of Normality of errors is not verified.
        \item For asymmetric distributions, the authors should be careful not to show in tables or figures symmetric error bars that would yield results that are out of range (e.g., negative error rates).
        \item If error bars are reported in tables or plots, the authors should explain in the text how they were calculated and reference the corresponding figures or tables in the text.
    \end{itemize}

\item {\bf Experiments compute resources}
    \item[] Question: For each experiment, does the paper provide sufficient information on the computer resources (type of compute workers, memory, time of execution) needed to reproduce the experiments?
    \item[] Answer: \answerYes{} %
    \item[] Justification: See the Efficiency paragraph in Appendix~\ref{app:additional_ablations}.
    \item[] Guidelines:
    \begin{itemize}
        \item The answer \answerNA{} means that the paper does not include experiments.
        \item The paper should indicate the type of compute workers CPU or GPU, internal cluster, or cloud provider, including relevant memory and storage.
        \item The paper should provide the amount of compute required for each of the individual experimental runs as well as estimate the total compute. 
        \item The paper should disclose whether the full research project required more compute than the experiments reported in the paper (e.g., preliminary or failed experiments that didn't make it into the paper). 
    \end{itemize}
    
\item {\bf Code of ethics}
    \item[] Question: Does the research conducted in the paper conform, in every respect, with the NeurIPS Code of Ethics \url{https://neurips.cc/public/EthicsGuidelines}?
    \item[] Answer: \answerYes{} %
    \item[] Justification: The research conforms to the NeurIPS Code of Ethics.
    \item[] Guidelines:
    \begin{itemize}
        \item The answer \answerNA{} means that the authors have not reviewed the NeurIPS Code of Ethics.
        \item If the authors answer \answerNo, they should explain the special circumstances that require a deviation from the Code of Ethics.
        \item The authors should make sure to preserve anonymity (e.g., if there is a special consideration due to laws or regulations in their jurisdiction).
    \end{itemize}

\item {\bf Broader impacts}
    \item[] Question: Does the paper discuss both potential positive societal impacts and negative societal impacts of the work performed?
    \item[] Answer: \answerYes{} %
    \item[] Justification: Broader impacts are discussed in Appendix~\ref{app:disc}.
    \item[] Guidelines:
    \begin{itemize}
        \item The answer \answerNA{} means that there is no societal impact of the work performed.
        \item If the authors answer \answerNA{} or \answerNo, they should explain why their work has no societal impact or why the paper does not address societal impact.
        \item Examples of negative societal impacts include potential malicious or unintended uses (e.g., disinformation, generating fake profiles, surveillance), fairness considerations (e.g., deployment of technologies that could make decisions that unfairly impact specific groups), privacy considerations, and security considerations.
        \item The conference expects that many papers will be foundational research and not tied to particular applications, let alone deployments. However, if there is a direct path to any negative applications, the authors should point it out. For example, it is legitimate to point out that an improvement in the quality of generative models could be used to generate Deepfakes for disinformation. On the other hand, it is not needed to point out that a generic algorithm for optimizing neural networks could enable people to train models that generate Deepfakes faster.
        \item The authors should consider possible harms that could arise when the technology is being used as intended and functioning correctly, harms that could arise when the technology is being used as intended but gives incorrect results, and harms following from (intentional or unintentional) misuse of the technology.
        \item If there are negative societal impacts, the authors could also discuss possible mitigation strategies (e.g., gated release of models, providing defenses in addition to attacks, mechanisms for monitoring misuse, mechanisms to monitor how a system learns from feedback over time, improving the efficiency and accessibility of ML).
    \end{itemize}
    
\item {\bf Safeguards}
    \item[] Question: Does the paper describe safeguards that have been put in place for responsible release of data or models that have a high risk for misuse (e.g., pre-trained language models, image generators, or scraped datasets)?
    \item[] Answer: \answerNA{} %
    \item[] Justification: No high-risk assets are released.
    \item[] Guidelines:
    \begin{itemize}
        \item The answer \answerNA{} means that the paper poses no such risks.
        \item Released models that have a high risk for misuse or dual-use should be released with necessary safeguards to allow for controlled use of the model, for example by requiring that users adhere to usage guidelines or restrictions to access the model or implementing safety filters. 
        \item Datasets that have been scraped from the Internet could pose safety risks. The authors should describe how they avoided releasing unsafe images.
        \item We recognize that providing effective safeguards is challenging, and many papers do not require this, but we encourage authors to take this into account and make a best faith effort.
    \end{itemize}

\item {\bf Licenses for existing assets}
    \item[] Question: Are the creators or original owners of assets (e.g., code, data, models), used in the paper, properly credited and are the license and terms of use explicitly mentioned and properly respected?
    \item[] Answer: \answerYes{} %
    \item[] Justification: All used public benchmarks and models are properly cited in the paper.
    \item[] Guidelines:
    \begin{itemize}
        \item The answer \answerNA{} means that the paper does not use existing assets.
        \item The authors should cite the original paper that produced the code package or dataset.
        \item The authors should state which version of the asset is used and, if possible, include a URL.
        \item The name of the license (e.g., CC-BY 4.0) should be included for each asset.
        \item For scraped data from a particular source (e.g., website), the copyright and terms of service of that source should be provided.
        \item If assets are released, the license, copyright information, and terms of use in the package should be provided. For popular datasets, \url{paperswithcode.com/datasets} has curated licenses for some datasets. Their licensing guide can help determine the license of a dataset.
        \item For existing datasets that are re-packaged, both the original license and the license of the derived asset (if it has changed) should be provided.
        \item If this information is not available online, the authors are encouraged to reach out to the asset's creators.
    \end{itemize}

\item {\bf New assets}
    \item[] Question: Are new assets introduced in the paper well documented and is the documentation provided alongside the assets?
    \item[] Answer: \answerYes{} %
    \item[] Justification: New assets introduced in the paper are well documented and the documentation is provided alongside the assets.
    \item[] Guidelines:
    \begin{itemize}
        \item The answer \answerNA{} means that the paper does not release new assets.
        \item Researchers should communicate the details of the dataset\slash code\slash model as part of their submissions via structured templates. This includes details about training, license, limitations, etc. 
        \item The paper should discuss whether and how consent was obtained from people whose asset is used.
        \item At submission time, remember to anonymize your assets (if applicable). You can either create an anonymized URL or include an anonymized zip file.
    \end{itemize}

\item {\bf Crowdsourcing and research with human subjects}
    \item[] Question: For crowdsourcing experiments and research with human subjects, does the paper include the full text of instructions given to participants and screenshots, if applicable, as well as details about compensation (if any)? 
    \item[] Answer: \answerNA{} %
    \item[] Justification: The paper does not involve crowdsourcing or human subjects.
    \item[] Guidelines:
    \begin{itemize}
        \item The answer \answerNA{} means that the paper does not involve crowdsourcing nor research with human subjects.
        \item Including this information in the supplemental material is fine, but if the main contribution of the paper involves human subjects, then as much detail as possible should be included in the main paper. 
        \item According to the NeurIPS Code of Ethics, workers involved in data collection, curation, or other labor should be paid at least the minimum wage in the country of the data collector. 
    \end{itemize}

\item {\bf Institutional review board (IRB) approvals or equivalent for research with human subjects}
    \item[] Question: Does the paper describe potential risks incurred by study participants, whether such risks were disclosed to the subjects, and whether Institutional Review Board (IRB) approvals (or an equivalent approval/review based on the requirements of your country or institution) were obtained?
    \item[] Answer: \answerNA{} %
    \item[] Justification: The paper does not involve human-subject research.
    \item[] Guidelines:
    \begin{itemize}
        \item The answer \answerNA{} means that the paper does not involve crowdsourcing nor research with human subjects.
        \item Depending on the country in which research is conducted, IRB approval (or equivalent) may be required for any human subjects research. If you obtained IRB approval, you should clearly state this in the paper. 
        \item We recognize that the procedures for this may vary significantly between institutions and locations, and we expect authors to adhere to the NeurIPS Code of Ethics and the guidelines for their institution. 
        \item For initial submissions, do not include any information that would break anonymity (if applicable), such as the institution conducting the review.
    \end{itemize}

\item {\bf Declaration of LLM usage}
    \item[] Question: Does the paper describe the usage of LLMs if it is an important, original, or non-standard component of the core methods in this research? Note that if the LLM is used only for writing, editing, or formatting purposes and does \emph{not} impact the core methodology, scientific rigor, or originality of the research, declaration is not required.
    \item[] Answer: \answerNA{} %
    \item[] Justification: LLMs are used only as backbones, baselines, or automatic judges, not as a non-standard core methodological component.
    \item[] Guidelines:
    \begin{itemize}
        \item The answer \answerNA{} means that the core method development in this research does not involve LLMs as any important, original, or non-standard components.
        \item Please refer to our LLM policy in the NeurIPS handbook for what should or should not be described.
    \end{itemize}

\end{enumerate}